\documentstyle[twoside,12pt,epsf,aps]{revtex}
 \relax
 \newcommand{\gevc} { {\rm GeV/c}}
 \newcommand{\gevcc}{ {\rm GeV/c^2}}
 \begin{document}
 \bibliographystyle{unsrt}
 \vskip 1cm
 \title{Measurement of the average time-integrated mixing probability of
       $b$-flavored hadrons produced at the Tevatron}
 \maketitle
 \font\eightit=cmti8
\def\r#1{\ignorespaces $^{#1}$}
\hfilneg
\begin{sloppypar}
\noindent
D.~Acosta,\r {14} T.~Affolder,\r 7 H.~Akimoto,\r {51}
M.G.~Albrow,\r {13} D.~Ambrose,\r {37}   
D.~Amidei,\r {28} K.~Anikeev,\r {27} J.~Antos,\r 1 
G.~Apollinari,\r {13} T.~Arisawa,\r {51}
% ...
\end{sloppypar}
\vskip .026in
\begin{center}
(CDF Collaboration)
\end{center}

\vskip .026in
\begin{center}
\r 1  {\eightit Institute of Physics, Academia Sinica, Taipei, Taiwan 11529, 
Republic of China} \\
\r 2  {\eightit Argonne National Laboratory, Argonne, Illinois 60439} \\
\r 3  {\eightit Istituto Nazionale di Fisica Nucleare, University of Bologna,
I-40127 Bologna, Italy} \\
\r 4  {\eightit Brandeis University, Waltham, Massachusetts 02254} \\
\r 5  {\eightit University of California at Davis, Davis, California  95616} \\
\r 6  {\eightit University of California at Los Angeles, Los 
Angeles, California  90024} \\ 
\r 7  {\eightit University of California at Santa Barbara, Santa Barbara, California 
93106} \\ 
\r 8 {\eightit Instituto de Fisica de Cantabria, CSIC-University of Cantabria, 
39005 Santander, Spain} \\
\r 9  {\eightit Carnegie Mellon University, Pittsburgh, Pennsylvania  15213} \\
\r {10} {\eightit Enrico Fermi Institute, University of Chicago, Chicago, 
Illinois 60637} \\
\r {11}  {\eightit Joint Institute for Nuclear Research, RU-141980 Dubna, Russia}
\\
\r {12} {\eightit Duke University, Durham, North Carolina  27708} \\
\r {13} {\eightit Fermi National Accelerator Laboratory, Batavia, Illinois 
60510} \\
\r {14} {\eightit University of Florida, Gainesville, Florida  32611} \\
\r {15} {\eightit Laboratori Nazionali di Frascati, Istituto Nazionale di Fisica
               Nucleare, I-00044 Frascati, Italy} \\
\r {16} {\eightit University of Geneva, CH-1211 Geneva 4, Switzerland} \\
\r {17} {\eightit Glasgow University, Glasgow G12 8QQ, United Kingdom}\\
\r {18} {\eightit Harvard University, Cambridge, Massachusetts 02138} \\
\r {19} {\eightit Hiroshima University, Higashi-Hiroshima 724, Japan} \\
\r {20} {\eightit University of Illinois, Urbana, Illinois 61801} \\
\r {21} {\eightit The Johns Hopkins University, Baltimore, Maryland 21218} \\
\r {22} {\eightit Institut f\"{u}r Experimentelle Kernphysik, 
Universit\"{a}t Karlsruhe, 76128 Karlsruhe, Germany} \\
\r {23} {\eightit High Energy Accelerator Research Organization (KEK), Tsukuba, 
Ibaraki 305, Japan} \\
\r {24} {\eightit Center for High Energy Physics: Kyungpook National
University, Taegu 702-701; Seoul National University, Seoul 151-742; and
SungKyunKwan University, Suwon 440-746; Korea} \\
\r {25} {\eightit Ernest Orlando Lawrence Berkeley National Laboratory, 
Berkeley, California 94720} \\
\r {26} {\eightit University College London, London WC1E 6BT, United Kingdom} \\
\r {27} {\eightit Massachusetts Institute of Technology, Cambridge,
Massachusetts  02139} \\   
\r {28} {\eightit University of Michigan, Ann Arbor, Michigan 48109} \\
\r {29} {\eightit Michigan State University, East Lansing, Michigan  48824} \\
\r {30} {\eightit Institution for Theoretical and Experimental Physics, ITEP,
Moscow 117259, Russia} \\
\r {31} {\eightit University of New Mexico, Albuquerque, New Mexico 87131} \\
\r {32} {\eightit Northwestern University, Evanston, Illinois  60208} \\
\r {33} {\eightit The Ohio State University, Columbus, Ohio  43210} \\
\r {34} {\eightit Osaka City University, Osaka 588, Japan} \\
\r {35} {\eightit University of Oxford, Oxford OX1 3RH, United Kingdom} \\
\r {36} {\eightit Universita di Padova, Istituto Nazionale di Fisica 
          Nucleare, Sezione di Padova, I-35131 Padova, Italy} \\
\r {37} {\eightit University of Pennsylvania, Philadelphia, 
        Pennsylvania 19104} \\   
\r {38} {\eightit Istituto Nazionale di Fisica Nucleare, University and Scuola
               Normale Superiore of Pisa, I-56100 Pisa, Italy} \\
\r {39} {\eightit University of Pittsburgh, Pittsburgh, Pennsylvania 15260} \\
\r {40} {\eightit Purdue University, West Lafayette, Indiana 47907} \\
\r {41} {\eightit University of Rochester, Rochester, New York 14627} \\
\r {42} {\eightit Rockefeller University, New York, New York 10021} \\
\r {43} {\eightit Instituto Nazionale de Fisica Nucleare, Sezione di Roma,
University di Roma I, ``La Sapienza," I-00185 Roma, Italy}\\
\r {44} {\eightit Rutgers University, Piscataway, New Jersey 08855} \\
\r {45} {\eightit Texas A\&M University, College Station, Texas 77843} \\
\r {46} {\eightit Texas Tech University, Lubbock, Texas 79409} \\
\r {47} {\eightit Institute of Particle Physics, University of Toronto, Toronto
M5S 1A7, Canada} \\
\r {48} {\eightit Istituto Nazionale di Fisica Nucleare, University of Trieste/\
Udine, Italy} \\
\r {49} {\eightit University of Tsukuba, Tsukuba, Ibaraki 305, Japan} \\
\r {50} {\eightit Tufts University, Medford, Massachusetts 02155} \\
\r {51} {\eightit Waseda University, Tokyo 169, Japan} \\
\r {52} {\eightit University of Wisconsin, Madison, Wisconsin 53706} \\
\r {53} {\eightit Yale University, New Haven, Connecticut 06520} \\
\end{center}

 \newpage
 \begin{abstract}
   We have measured the number of like-sign (LS) and opposite-sign (OS) 
   lepton pairs arising from double semileptonic decays of $b$ and  
   $\bar{b}$-hadrons, pair-produced at the Fermilab Tevatron collider. 
   The data samples were collected with the Collider Detector at Fermilab
   (CDF) during the $1992-1995$ collider run by triggering on the existence
   of $\mu \mu$ or $e\mu$ candidates in an event. The observed ratio of LS
   to OS dileptons leads to a measurement of the average time-integrated 
   mixing probability of all produced $b$-flavored hadrons which decay weakly, 
   $\bar{\chi}= 0.152 \pm 0.007\; ({\rm stat.}) \pm 0.011\; ({\rm syst.})$,
   that is significantly larger than the world average   
   $\bar{\chi} = 0.118 \pm 0.005$.\\
   PACS number(s): 13.85.Qk, 13.20.Jf
 \end{abstract}
 \section {Introduction}
 \label{sec:ss-intro}
   The time evolution of $B_d^0 - \bar{B}_d^0$ mixing has been accurately 
   measured in a number of experiments, while $B_s^0 - \bar{B}_s^0$ mixing
   has not yet been observed. Time-independent measurements of $B^0$ mixing
   offer an experimentally distinct technique to extract $B^0$ mixing
   parameters. The time-integrated mixing probability is defined as 
   $\bar{\chi} = \frac{ \Gamma (B^0 \to \bar{B}^0 \to \ell^+X)} 
                      {\Gamma (B \to \ell^{\pm}X)} $,
   where the numerator includes $B_d^0$ and $B_s^0$ mesons and the 
   denominator includes all $B$ hadrons. The average probability is then
   $\bar{\chi} = f_d \cdot \chi_d + f_s \cdot \chi_s$, where $\chi_d$ and 
   $f_d$, and $\chi_s$ and $f_s$ are the time-integrated mixing probability
   and the fraction of produced $B_d^0$ and $B_s^0$ mesons, respectively, that
   decay semileptonically. A measurement of $\bar{\chi}$ can be used to
   extract $B^0$ mixing information through $\chi_d$ and $\chi_s$, or,
   alternatively, to extract information on the fractions of produced $B_d^0$ 
   and $B_s^0$ mesons. 

   A precise measurement of the time-integrated mixing probability $\bar{\chi}$
   at the Tevatron can also provide indications for new physics through its
   comparison with the LEP measurements and the time-dependent results from
   the Tevatron. For example, a recent publication~\cite{berger} explores an
   explanation within the context of the minimal supersymmetric standard model
   for the long-standing discrepancy between the measured cross section for
   bottom-quark production at the Tevatron and the next-to-leading order (NLO)
   prediction. Ref.~\cite{berger} postulates the existence of a relatively
   light gluino $\tilde{g}$ (mass $\simeq$ 12 to 16 GeV/$c^2$) that decays
   into a $b$ quark and a light $\tilde{b}$ squark (mass $\simeq$ 2 to 
   5.5 GeV/$c^2$). The pair production of such light gluinos provides a
   bottom-quark cross section comparable in magnitude to the conventional-QCD
   component. Since $\tilde{g}$ is a Majorana particle, its decay yields both
   quark and antiquark; therefore, gluino pair production and subsequent decay
   to $b$-quarks will generate $bb$ and $\bar{b}\bar{b}$ pairs, as well as the
   $b\bar{b}$ final states that appear in conventional QCD production. 
   The pair production of gluinos leads therefore to an increase of like-sign
   dileptons from weak decays of $b$ quarks~\footnote{Constraints to this
   scenario have been derived from other data analyses (see, for example,
   Ref.~\cite{rosner} and experimental references therein).}.
   This increase could be confused with an enhanced rate of $B^0 -\bar{B}^0$ 
   mixing and result in a value of $\bar{\chi}$ larger than the world average
   $0.118 \pm 0.005$~\cite{pdg}. Using a previous CDF result~\cite{yale} 
   ($\bar{\chi}= 0.131 \pm 0.020\; ({\rm stat.}) \pm 0.016\; ({\rm syst.})$,
   Ref.~\cite{berger} estimates that the value of $\bar{\chi}$ at the Tevatron
   could be as large 0.17~\footnote{
   Determinations of $\chi_d$~\cite{timemx}, based on the direct measurement
   of the oscillation frequency $\Delta m_d$, are not sensitive to this type 
   of unconventional $b\bar{b}$ production; in fact, an extra source of
   like-sign $b$ quarks, would reduce the amplitude of the mixing asymmetry,
   but would not affect the determination of $\Delta m_d$.}.
   The $\bar{\chi}$ measurement in Ref.~\cite{yale} is based upon muon pairs
   corresponding to an integrated luminosity of 17.4 pb$^{-1}$. The present 
   measurement, which makes use of a dimuon data set corresponding to an 
   integrated luminosity of 105 pb$^{-1}$ and an $e \mu$ data set 
   corresponding to approximately 85 pb$^{-1}$, supersedes our previous result.

   In this study, the time-integrated mixing probability $\bar{\chi}$ is 
   derived from the ratio of the observed numbers of LS and OS lepton pairs
   arising from $b\bar{b}$ production. At the Tevatron, dilepton events  
   result from decays of heavy quark pairs ($b\bar{b}$ and $c\bar{c}$),
   the Drell-Yan process, charmonium and bottomonium decays, and decays of
   $\pi$ and $K$ mesons. Background to dilepton events also comes from the 
   misidentification of $\pi$ or $K$ mesons. As in Ref.~\cite{yale}, we make
   use of the precision tracking provided by the CDF silicon microvertex
   detector to evaluate the fractions of leptons due to long-lived $b$- and
   $c$-hadron decays, and to the other background contributions.

   Sections~\ref{sec:ss-det} and~\ref{sec:ss-anal} describe the detector
   systems relevant to this analysis and the data selection, respectively.
   The analysis method, similar to the one used in Ref.~\cite{yale}, is
   discussed in Sec.~\ref{sec:ss-meth}. In Sec.~\ref{sec:ss-res}, we determine
   the contributions of the $b\bar{b}$ and $c\bar{c}$ production to OS and LS
   dileptons. The $B^0 -\bar{B}^0$ mixing result is derived in 
   Sec.~\ref{sec:ss-chib}. Section~\ref{sec:ss-cross} presents cross-checks
   and studies of systematics effects. Our conclusions are
   summarized in Sec.~\ref{sec:ss-concl}.
 \section{CDF detector and trigger}
 \label{sec:ss-det}
   The CDF detector is described in detail in Ref.~\cite{cdf-det}. We review
   the detector components most relevant to this analysis. Inside the 1.4 T
   solenoid the silicon microvertex detector (SVX)~\cite{svx}, a vertex drift
   chamber (VTX), and the central tracking chamber (CTC) provide the tracking
   and momentum information for charged particles. The CTC is a cylindrical 
   drift chamber containing 84 measurement layers. It covers the 
   pseudorapidity interval $|\eta|\leq 1.1$, where 
   $\eta =- \ln [\tan (\theta/2)]$. In CDF, $\theta$ is the polar angle 
   measured from the proton direction, $\phi$ is the azimuthal angle, and $r$
   is the radius from the beam axis ($z$-axis). The SVX consists of four
   layers of silicon micro-strip detectors located at radii between 2.9 and
   7.9 cm from the beam line and provides spatial measurements in the $r-\phi$
   plane with a resolution of 13 $\mu$m. It gives a track impact 
   parameter~\footnote{The impact parameter is the distance of closest
   approach of a track to the primary event vertex in the transverse plane.}
   resolution of about $(13+40/p_{T})$ $\mu$m, where $p_T$ is the track 
   momentum measured in the plane transverse to the beam axis and in
   $\gevc$ units. The SVX extends $\pm 25$ cm along the $z$-axis. Since the
   vertex $z$-distribution for $p\bar{p}$ collision is approximately a
   Gaussian function with an rms width of 30 cm, the average geometric 
   acceptance of the SVX is about 60\%. The transverse profile of the Tevatron
   beam is circular and has an rms spread of $\simeq 30\; \mu$m in the
   horizontal and vertical directions. The $p_T$ resolution of the combined
   CTC and SVX detectors is $\delta p_T/p_T=[(0.0066)^2+(0.0009\; (\gevc)^{-1}
   \cdot p_T )^2]^{1/2}$. Electromagnetic (CEM) and hadronic (CHA) calorimeters
   with projective tower geometry are located outside the solenoid and cover 
   the pseudorapidity region $|\eta|\leq 1.1$, with a segmentation of
   $\Delta \phi =15^{\deg}$ and $\Delta \eta=0.11$. A layer of proportional
   chambers (CES) is embedded near shower maximum in the CEM and provides a
   more precise measurement of the electromagnetic shower position. Two muon
   subsystems in the central rapidity region ($|\eta|\leq 0.6$) are used for
   muon identification: the central muon chambers (CMU), located behind the
   CHA calorimeter, and the central upgrade muon chambers (CMP), located
   behind an additional 60 cm of steel.

   CDF uses a three-level trigger system. At the first two levels, decisions
   are made with dedicated hardware. The information available at this stage
   includes energy deposit in the CEM and CHA calorimeters, high-$p_T$ tracks
   found in the CTC by a fast track processor, and track segments found in the
   muon subsystems. At the third level of the trigger, events are selected
   based on a version of the off-line reconstruction programs optimized for 
   speed. The lepton selection criteria used by the 3rd level trigger are
   similar to those described in the next section.

   A large fraction of the events used for this analysis are collected using
   two triggers that require two lepton candidates in an event. The first
   trigger requires two muon candidates; each muon candidate requires a track
   in the CTC, matched with track segments in the CMU system, corresponding to
   a particle with $p_T \geq 2.2 \; \gevc$. At least one of the candidates is
   required to have track segments in both the CMU and CMP chambers. The
   second trigger requires an electron and a muon candidate. The $E_T$ 
   threshold for the electron is 5 GeV, where $E_T = E \sin \theta$, and $E$
   is the energy measured in the CEM. In addition, the trigger requires the
   presence of a CTC track with $p_T \geq 4.7 \; \gevc$ and the same $\phi$
   angle of the CEM energy deposit. The muon candidate requires a CTC track
   with matched segments in the CMU chambers and $p_T \geq 2.7 \; \gevc$. 
 \section{Data selection}
 \label{sec:ss-anal}
   For this analysis we select events which contain two and only two good
   leptons. Good muons are selected by requiring $p_T \geq 3 \; \gevc$
   and a match between the CTC track extrapolated in the muon chambers
   and the muon segment within $3\; \sigma$ in the $r-\phi$ plane (CMU and
   CMP) and $\sqrt{12}\; \sigma$ in the $r-z$ plane (CMU), where $\sigma$ is
   a standard deviation including the effect of multiple scattering. In order
   to minimize misidentification of muons due to hadronic punchthrough, we 
   require a muon segment in the CMP chambers as well as an energy deposit in
   the calorimeters larger than 0.1 GeV but smaller than 2 and 6 GeV in the
   CEM and CHA, respectively. The identification of good electrons makes use
   of the information from calorimeters and tracking chambers. We select 
   electrons with $E_T \geq 5$ GeV, and, as in previous analyses~\cite{topxs},
   we require the following:
   (1) the ratio of hadronic to electromagnetic energy of the cluster,
   $E_{had}/E_{em} \leq 0.05 $; (2) the ratio of cluster energy to track
   momentum, $E/P \leq 1.5$; (3) a comparison of the lateral shower profile
   in the calorimeter cluster with that of test-beam electrons, 
   $L_{shr} \leq 0.2 $; (4) the distance between the extrapolated 
   track-position and the CES measurement in the $r-\phi$ and $z$ views, 
   $\Delta x \leq 1.5\; {\rm cm}$ and $\Delta z \leq 3.0\; {\rm cm}$;
   (5) a $\chi^{2}$ comparison of the CES shower profile with those of 
   test-beam electrons, $\chi^{2}_{strip} \leq 15$. Fiducial cuts on the 
   electromagnetic shower position as measured in the CES, are applied to 
   ensure that the electron candidate is away from the calorimeter boundaries
   and the energy is well measured. Electrons from photon conversions are 
   removed using an algorithm based on track information~\cite{topxs}. 

   To ensure accurate impact parameter measurement, each lepton track is
   required to be reconstructed in the SVX with hits non-shared with other
   tracks in at least two layers out of the possible four. We also require
   the impact parameter of each lepton track to be less than 0.2 cm with
   respect to the primary vertex~\footnote{This cut removes most of the cosmic
   rays, since this background is distributed as a linear function of the
   impact parameter.}.
   Lepton tracks are required to be within 5 cm from the primary vertex in the
   $z$-direction. To reconstruct the primary event vertex, we first identify
   its $z$-position using the tracks reconstructed in the VTX detector. When
   projected back to the beam axis, these tracks determine the longitudinal
   position with a precision of about 0.2 cm. The transverse position of the
   primary vertex is determined for each event by a weighted fit of all SVX
   tracks which have a $z$ coordinate within 5 cm of the $z$-vertex position
   of the primary vertex. First, all tracks are constrained to originate
   from a common vertex. The position of this vertex is constrained by the 
   transverse beam envelope described above. Tracks that have impact 
   parameter significance $|d|/\sigma_{d}$, where $\sigma_{d}$ is the estimate
   of the uncertainty on the impact parameter $d$, larger than three with 
   respect to this vertex are removed and the fit is repeated. This procedure
   is iterated until all used tracks satisfy the impact parameter requirement.
   At least five tracks must be used in the determination of the transverse
   position of the primary vertex or we use the nominal beam-line position. 
   We use this procedure to avoid having the primary vertex position biased by
   the presence of heavy flavor decays~\cite{topxs}. The primary vertex
   coordinates transverse to the beam direction have uncertainties in the
   range of 10$-$25 $\mu$m, depending on the number of tracks and the event
   topology.
 
   In the analysis, lepton pairs arising from $b$ cascade decays are removed 
   by selecting dilepton candidates with invariant mass greater than 
   5 GeV/c$^2$.
  \section{Method of analysis}
  \label{sec:ss-meth}

   For leptons originating from the decay of long lived particles the impact
   parameter is $d=|\beta \gamma c t \sin(\delta)|$, where $t$ is the proper
   decay time of the parent particle from which the lepton track originates,
   $\delta$ is the decay angle of the lepton track with respect to the
   direction of the parent particle, and $\beta \gamma$ is a Lorentz boost
   factor. The impact parameter of the lepton is proportional to the lifetime
   of the parent particle. The markedly different impact parameter 
   distributions for leptons from $b$ decays, $c$ decays, and other sources
   allow the determination of the parent fractions.
  
   The method used to determine the $b\bar{b}$ and $c\bar{c}$ content of the
   data has been pioneered in Ref.~\cite{yale}. The procedure is to fit the
   observed impact parameter distribution of the lepton pairs with the
   expected impact parameter distributions of leptons from various sources.
   After data selection, the main sources of reconstructed leptons are
   semileptonic decays of bottom and charmed hadrons, and prompt decays
   of onia and Drell-Yan production.

\begin{figure}
 \begin{center}
 \vspace{-0.3in}
 \leavevmode
 \epsffile{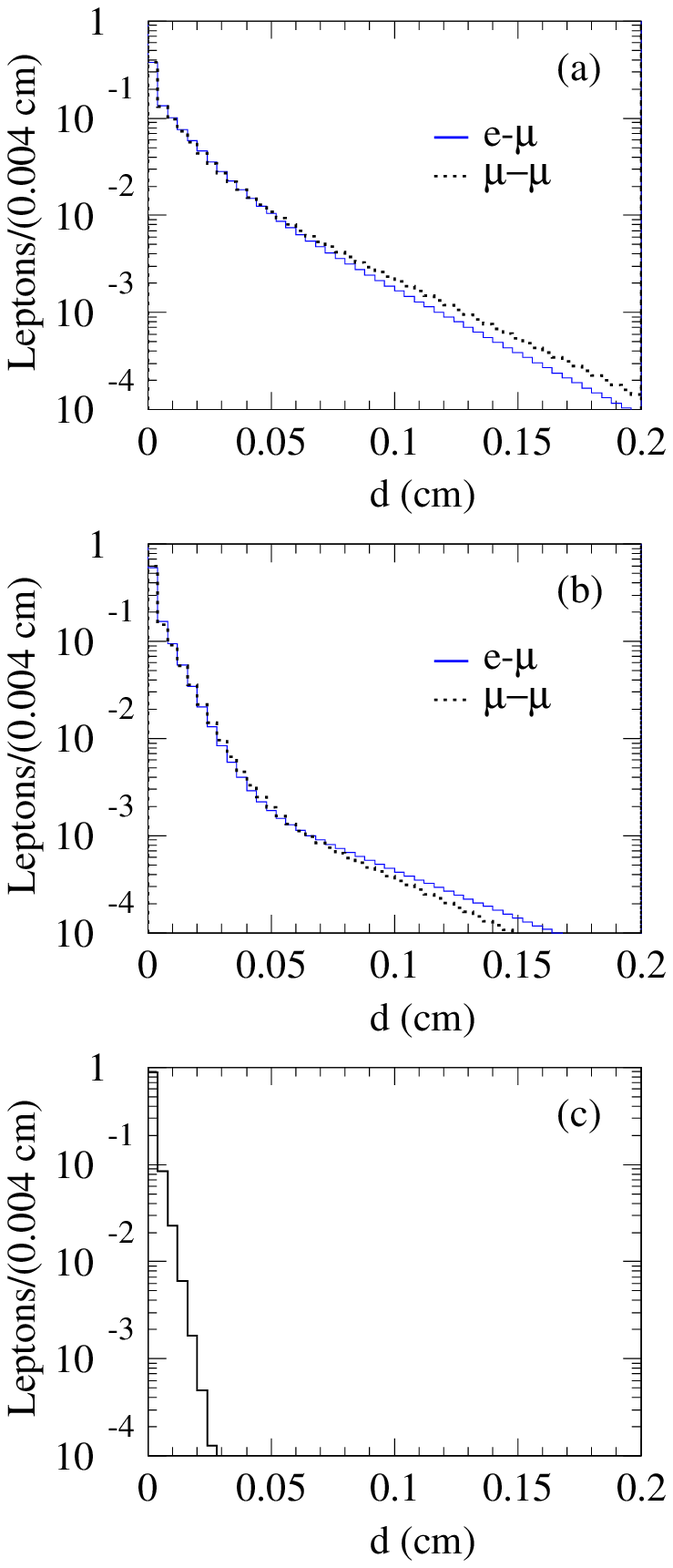}
 \caption[]{Impact parameter distributions of leptons coming from 
            $b$ decays (a), $c$ decays (b), and prompt leptons (c).
	    Distributions are normalized to unit area; differences between
	    $\mu - \mu$ and $e - \mu$ templates are due to the different 
	    $p_T$ thresholds. The ratio of the number of events with
            $d \leq 0.008$ cm to that with $d \geq 0.008$ cm is 1.04, 2.85,
	    and 32.3 for the histograms (a), (b), and (c), respectively.}
 \label{fig:fig_2}
 \end{center}
 \end{figure}

   Monte Carlo simulations are used to model the impact parameter 
   distributions for leptons from $b$ and $c$ decays. We use the {\sc herwig}
   Monte Carlo generator program~\cite{herw} to generate hadrons with heavy
   flavors~\footnote{
   We use option 1500 of version 5.6, generic $2 \rightarrow 2$ hard
   scattering with $p_T \geq 5 \; \gevc$, with the same setting of the 
   {\sc herwig} parameters used in Ref.~\cite{topxs}. In the generic hard
   parton scattering, $b\bar{b}$ and $c\bar{c}$ pairs are generated by
   {\sc herwig} through processes of order $\alpha_s^{2}$ (LO) such as 
   $gg \rightarrow b\bar{b}$ (direct production). Processes of order
   $\alpha_s^{3}$  are implemented in {\sc herwig} through flavor excitation
   processes, such as $gb \rightarrow g b$, or gluon splitting, in which the
   process $gg \rightarrow gg$ is followed by $g \rightarrow b\bar{b}$.},
   the {\sc qq} Monte Carlo program~\cite{cleo} to decay hadrons with heavy
   flavor, and the {\sc qfl} Monte Carlo simulation of CDF~\cite{topxs} to
   model the detector's response. Impact parameter distributions for simulated
   $b$ and $c$ decays are shown in Figures~\ref{fig:fig_2}(a) and (b),
   respectively. Since lifetimes of bottom and charmed hadrons 
   ($c\tau_{B} \simeq 480\; \mu$m and $c\tau_{D} \simeq 200\; \mu$m) are much
   larger than the average SVX impact parameter resolution in these data sets
   ($\simeq 15\; \mu$m), the dominant factor determining the impact parameter
   distribution is the kinematics of the semileptonic decays which is well
   modeled by the simulation (see Sect.~\ref{sec:ss-cross}). The fraction of
   leptons from sequential $b$ decays ($b \rightarrow c X, c \rightarrow l Y$)
   is also determined with the simulation. Leptons from sequential $b$ decays
   have slightly different kinematics and slightly larger $ct$ than 
   leptons coming from direct $b$ decays; these two effects compensate and 
   the simulated impact parameter distribution of leptons from sequential
   decays is indistinguishable from that of leptons from direct $b$ decays. 

\begin{figure}[]
 \begin{center}
 \vspace{-0.2in}
 \leavevmode
 \epsffile{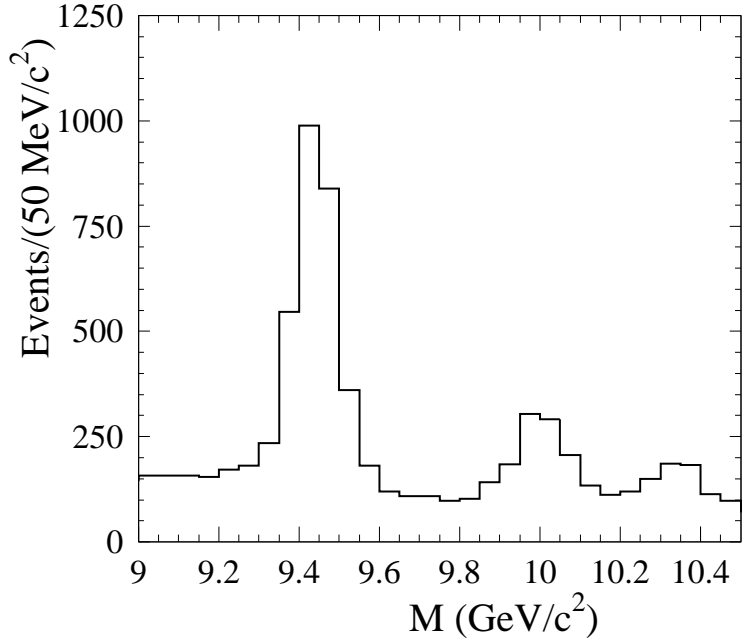}
 \caption[]{Invariant mass distribution of OS dimuons in the $\Upsilon$ region.
	    The impact parameter distribution in Figure~\ref{fig:fig_2}(c) is
	    derived using muons with invariant mass between 9.28 and 9.6
            GeV/$c^2$. The background is removed using dimuons with invariant
            mass between 9.04 and 9.2 GeV/$c^2$ and between 9.54 and 9.7 
            GeV/$c^2$. Dimuon events in the mass range $9.2- 10.5\; \gevcc$, 
            which are dominated by $\Upsilon$ production, are not used in the
            $\bar{\chi}$ analysis.}
 \label{fig:fig_1}
 \end{center}
 \end{figure}

   The impact parameter distribution of leptons from prompt sources such as
   quarkonia decays and Drell-Yan production is plotted in 
   Fig.~\ref{fig:fig_2}(c) and is derived using muons from $\Upsilon(1S)$ 
   decays~\footnote{
   We use templates derived from the data to account properly for non-Gaussian
   tails of the impact parameter distribution. The impact parameter 
   distribution of electrons from a smaller statistics sample of 
   $Z \rightarrow e^+ e^-$ is also well modeled by the muon template.}
   (see Fig.~\ref{fig:fig_1}).
 
   Lepton tracks from $\pi$ and $K$ in-flight decays are also regarded as 
   prompt tracks since the track reconstruction algorithm rejects tracks with
   appreciable kinks. Tracks of $\pi$ and $K$ mesons, which mimic the lepton
   signal, are also regarded as prompt since the average heavy flavor
   contribution per event is negligible (see Sect.~\ref{sec:ss-cross}).

   Since there are two leptons in an event, the fit is performed in the 
   two-dimensional space of impact parameters. Each axis represents the impact
   parameter of one of the two leptons. In filling the histograms, the lepton
   ordering by flavor type or transverse momentum is randomized. The 
   two-dimensional impact parameter technique exploits the fact that the
   lepton impact parameters are independent uncorrelated
   variables~\footnote{
   The correlation between the two impact parameters, $\rho=\frac{\int \int
   (d_1-<d_1>)(d_2-<d_2>) \delta d_1 \delta d_2}{\sigma_{d_1} \sigma_{d_2}}$, 
   is approximately 0.04 in the data samples and their heavy flavor 
   simulations.}. 
   The two-dimensional template distributions for each type of event are made
   by combining the relevant one-dimensional distributions in
   Fig.~\ref{fig:fig_2}.

\begin{figure}[htb]
 \begin{center}
 \vspace{-0.2in}
 \leavevmode
 \epsfxsize \textwidth
 \epsffile{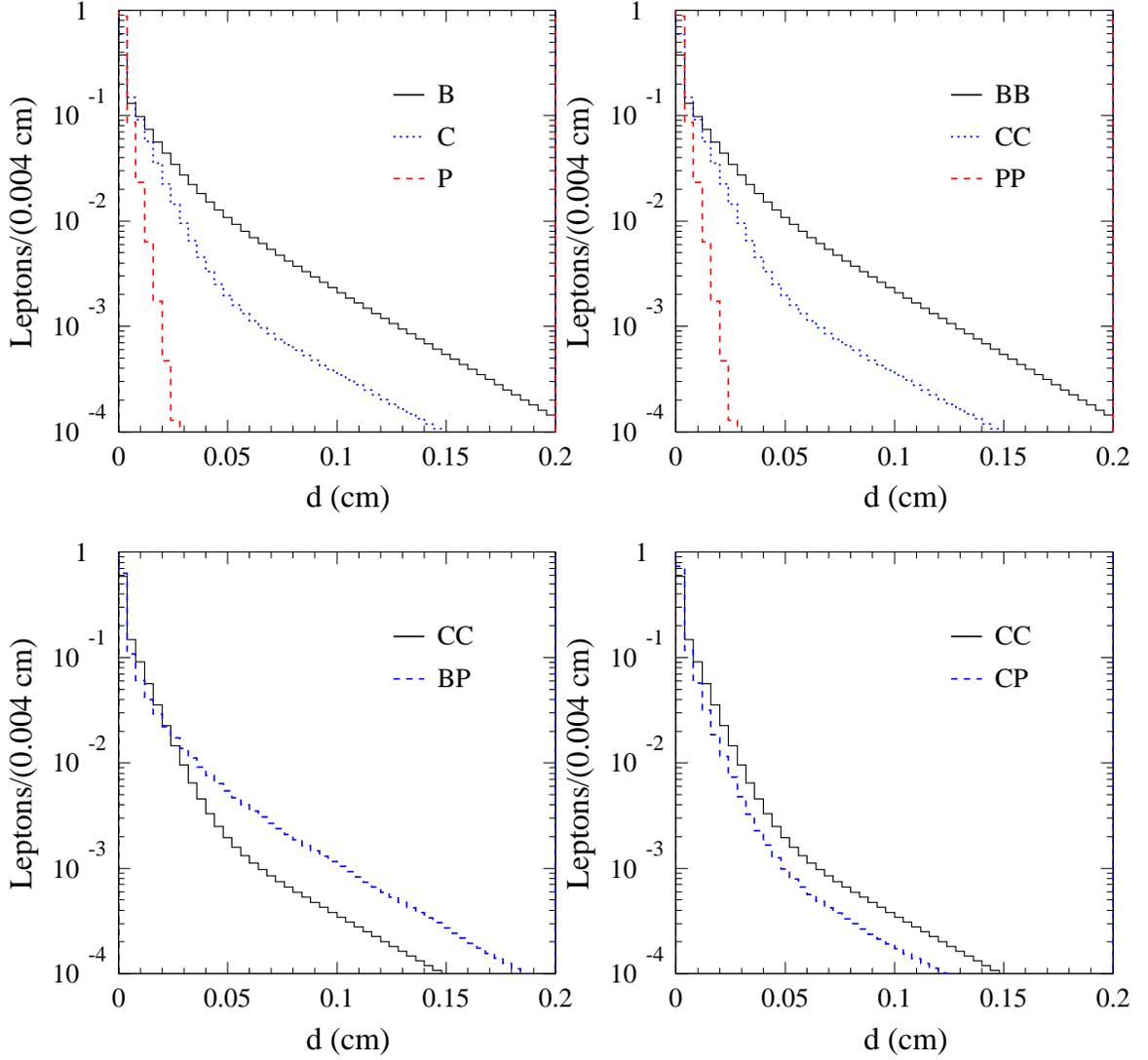}
 \caption[]{Projections of the two-dimensional impact parameter distributions 
            of the different components used to fit the dimuon data (see text).
            The top-left distribution shows the shapes of the prompt, $b$ and
            $c$ templates used to construct the different two-dimensional
            distributions used in the likelihood function. All distributions 
            are normalized to unit area.}
 \label{fig:fig_comp}
 \end{center}
 \end{figure}

   A binned maximum log likelihood method is used to fit simultaneously the
   impact parameter distributions of OS and LS dileptons. The likelihood $L$
   is defined as
   \begin{eqnarray*}
      L = \prod_i \prod_j [ l_{ij}^{n(i,j)} e^{-l_{ij}}/n(i,j)!] 
   \end{eqnarray*}
   where $n(i,j)$ is the number of events in the $(i,j)$th bin. The function 
   $l_{ij}$ is defined as
   \begin{eqnarray*}
    l_{ij} &  = & BB \cdot S_b(i)\cdot S_b(j) + CC \cdot S_c(i)\cdot S_c(j) 
                + PP \cdot S_p(i)\cdot S_p(j) + \\ 
    & &    0.5 \cdot [ BP \cdot (S_b(i)\cdot S_p(j) + S_p(i)\cdot S_b(j) ) +
     CP \cdot (S_c(i)\cdot S_p(j) + S_p(i)\cdot S_c(j) )]   
   \end{eqnarray*}
   where $S_b$, $S_c$, and $S_p$ are the impact parameter templates shown in
   Fig.~\ref{fig:fig_2}(a), (b), and (c), respectively. The fit parameters
   $BB$, $CC$, and $PP$ represent the $b\bar{b}$, $c\bar{c}$ and prompt 
   dilepton contributions, respectively. The fit parameter $BP$ ($CP$) 
   estimates the number of events in which there is only one $b$ ($c$) quark
   in the detector acceptance and the second lepton is produced by the decay
   or the misidentification of $\pi$ and $K$ mesons~\footnote{
   According to the simulation, supported by the measurement in Ref.~\cite{d0},
   approximately 90\% of the $b\bar{b}$ and $c\bar{c}$ events with an 
   identified lepton from heavy flavor decay do not contain the second heavy 
   flavored hadron in the detector acceptance. Therefore, we ignore the 
   small contribution to misidentified leptons due to $\pi$ and $K$ mesons 
   from heavy flavor decays (see Sec.~\ref{sec:ss-cross}).}.
   Figure~\ref{fig:fig_comp} compares projections of the two-dimensional 
   distributions for each type of dilepton contribution to the likelihood.
   Because of sequential decay and mixing, the $b\bar{b}$ production results
   in both OS and LS dileptons. For LS dileptons, one expects no contribution
   from $c\bar{c}$ production.

   We do not fit dimuon events with invariant mass between 9.2 and 10.5
   GeV/$c^2$ since OS dimuons are dominated by $\Upsilon$ meson production.
   The $PP$ contribution to $e \mu$ events can only arise from misidentified
   leptons ($\tau \tau$ Drell-Yan production is negligible)
   and is expected to be equal for OS and SS dileptons. Therefore, in the fit
   to $e \mu$ data, the $PP$ components in OS and LS dileptons are constrained
   to be equal within the statistical error (technically, we add the term
   $0.5 \times (PP(OS) - PP(LS))^2/((PP(OS)+PP(LS))$ to the function $-\ln L$ 
   used by the fit). In dimuon events, where the Drell-Yan contribution is
   relevant, OS leptons have a larger $PP$ component than LS dileptons. The 
   $BP$ and $CP$ contributions, in which one lepton is fake, are expected to
   be the same for OS and LS dileptons, and in the fit are constrained to be
   equal within the statistical error. One also expects the $BP$ and $CP$ 
   contributions to have approximately the same size~\footnote{
   According to the simulation, the cross section for producing at least one
   $c$ hadron in the detector acceptance is approximately a factor of two 
   larger than the cross section for producing at least one $b$ hadron in the
   detector acceptance. Since the efficiency for detecting a lepton from a $c$
   decay is approximately 40\% of that for detecting a lepton from a $b$ decay,
   one expects the $b\bar{b}$ and $c\bar{c}$ contributions to events with at
   least one identified lepton to be approximately equal. In contrast, the
   $b\bar{b}$ and $c\bar{c}$ cross sections for producing events which contain
   2 hadrons with heavy flavor in the detector acceptance are dominated by the
   LO term and are approximately equal; one therefore expects the $b\bar{b}$
   contribution to dilepton events to be much larger than the $c\bar{c}$ 
   contribution. }.
 \section{Result}
 \label{sec:ss-res}
   We show the result of the fit to the data for dimuon and $e\mu$ events
   in subsections~A and B, respectively.
 \subsection{Dimuon events}
 \label{sec:ss-dimu}
\begin{figure}[htb]
 \begin{center}
 \vspace{-0.2in}
 \leavevmode
 \epsfxsize \textwidth
 \epsffile{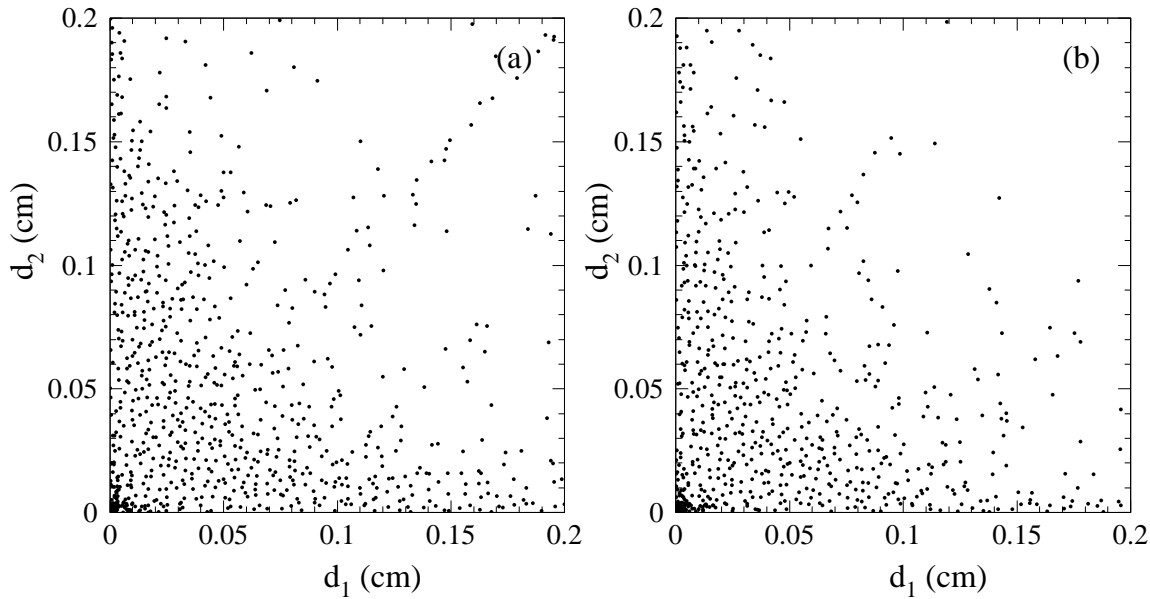}
 \caption[]{Two dimensional impact parameter distributions for (a) OS and 
	    (b) LS dimuons.}
 \label{fig:fig_3}
 \end{center}
 \end{figure}

    The observed two-dimensional impact parameter distributions for OS and LS
    dimuons are plotted in Figure~\ref{fig:fig_3}. We do not use dimuon events
    with invariant mass between 9.2 and 10.5 GeV/$c^2$ since OS are largely 
    dominated by $\Upsilon$ meson production.  There are 18420 OS dimuons and
    9279 LS dimuons after the removal of 6264 OS and 1302 LS dimuons with
    invariant mass in the $\Upsilon$ region.
    One sees that a handful of events in Fig.~\ref{fig:fig_3}(a) cluster 
    along the diagonal line $d_1 = d_2$. These events are due to cosmic rays.
    We minimize their contribution by fitting only events with 
    $d_1+d_2\leq 0.2$ cm. As shown in Sec.~\ref{sec:ss-cross}, the fit result
    is unaffected by the inclusion of events with  $d_1 +d_2 \geq 0.2$ cm.
    When all the likelihood terms are used to fit the data, the best fit,
    as expected, returns $CC=0 \pm 40$ LS events. However, while the fit 
    finds an appreciable $BP$ component, it returns $CP =0 \pm 110$ in both
    LS and OS events. When fitting the data with all components, the fit 
    gets blocked when limiting the $CC(LS)$ and $CP$ parameters to positive 
    values, and it returns reliable errors only when allowing the $CC$ and
    $CP$ terms to have also unphysical (negative) values. Since these
    unphysical values produce an overestimate of the size and the error of the
    remaining components, we fit again the data setting to zero the $CC$ term
    in LS events and the $CP$ contribution to OS and LS events~\footnote{
    In Sec~\ref{sec:ss-cross}, we show that this happens in 15\% of simulated
    pseudo-experiments due to the fact that $CC$, $BP$ and $CP$ templates are 
    quite similar. In addition, we show that the fit result does not vary when
    constraining the $BP$ and $CP$ components to be, as expected, equal
    within their statistical error.}. 

\begin{figure}[htb]
 \begin{center}
 \vspace{-0.2in}
 \leavevmode
 \epsfxsize \textwidth
 \epsffile{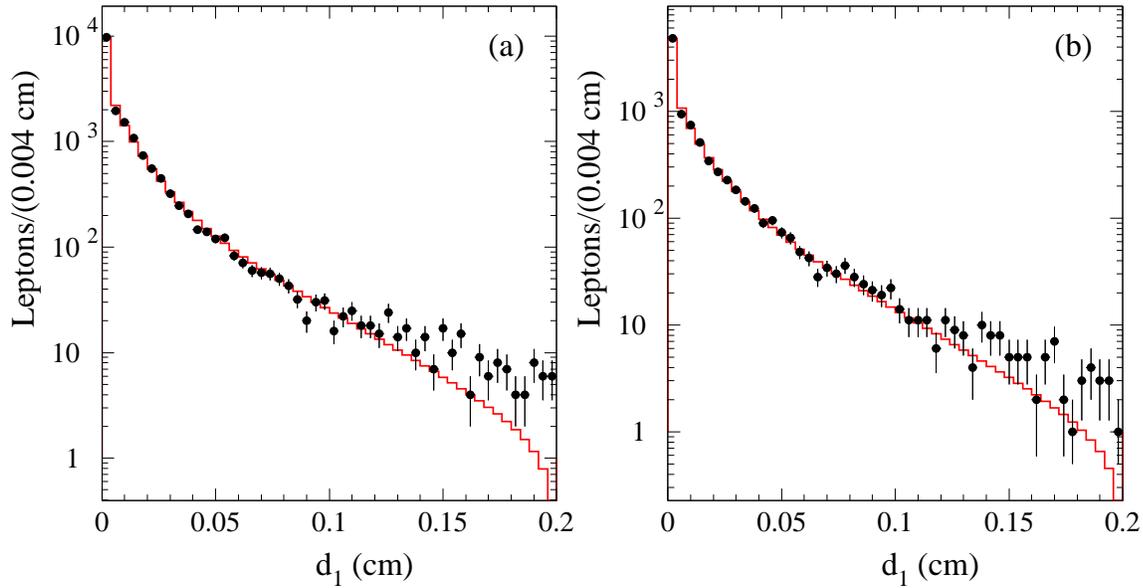}
 \caption[]{The projection of the impact parameter distribution of (a) OS
           and (b) LS dimuons onto one of the two axis is compared to the fit.}
 \label{fig:fig_4}
 \end{center}
 \end{figure}

    The fit result is shown in Table~\ref{tab:tab_1}. The parameter 
    correlation matrix is listed in Table~\ref{tab:corr_dimu}. The best fit 
    returns $-\ln L = 3076$. The probability of the $-\ln L$ value returned by
    the fit is determined by fitting Monte Carlo pseudoexperiments. In each
    experiment, we randomly generate different components with average size as
    determined by the fit to the data and allowing for Poisson fluctuations; 
    the impact parameter distribution for each component is randomly generated
    from the corresponding templates used in the fit. We find that 40\% of the
    fits to the pseudoexperiments return a $-\ln L$ value equal or larger than
    $3076$. For a comparison of the data and the fit results, projections of
    the two-dimensional impact parameter distributions are shown in
    Fig.~\ref{fig:fig_4}. Since the fit appears to underestimate the data for
    $d_1 \geq$ 0.12 cm, we have fitted the data excluding points at impact 
    parameters larger than 0.12 cm; this fit returns a result identical to
    that of the standard fit.

    Using Table~\ref{tab:tab_1}, one derives a ratio of LS to OS dimuons due 
    to $b\bar{b}$ production which is $R = 0.537 \pm 0.018$.  
 \begin{table}[htb]
 \begin{center}
 \caption[]{Number of events attributed to the different sources of dimuons by
           the fit to OS and LS dimuons with $d_1+d_2 \leq 0.2$ cm. 
           The errors correspond to a 0.5 change of the $-\ln L$.}
 \begin{tabular}{lcc}
 Component  &         OS      &        LS       \\
  $BB$      & $10476 \pm 223$ & $5630 \pm  132$ \\ 
  $CC$      & $~2469 \pm 360$ &    ~~~~0~~~     \\
  $PP$      & $~3603 \pm 161$ & $1914 \pm 87~$  \\
  $BP$      & $~1566 \pm 165$ & $1555 \pm 157$  \\
  $CP$      &    ~~~~~0~~~    & ~~~~0~~~     \\
 \end{tabular}
 \label{tab:tab_1}
 \end{center}
 \end{table}
 \begin{table}[htb]
 \begin{center}
 \caption[]{Parameter correlation coefficients returned by the fit listed
           in Table~\ref{tab:tab_1}. }
 \begin{tabular}{lcccccc}
  Component & $BB(OS)$ & $CC(OS)$ & $PP(OS)$ & $BP(OS)$ & $BB(LS)$ & $PP(LS)$\\
  $CC(OS)$  & $-0.70$  &          &          &          &          &         \\
  $PP(OS)$  & $~0.53$  & $-0.73$  &          &          &          &         \\
  $BP(OS)$  & $-0.03$  & $-0.46$  & $~0.05$  &          &          &         \\
  $BB(LS)$  & $~0.02$  & $~0.31$  & $-0.03$  & $-0.66$  &          &         \\
  $PP(LS)$  & $~0.02$  & $~0.27$  & $-0.03$  & $-0.58$  & $~0.25$  &         \\
  $BP(LS)$  & $-0.03$  & $-0.44$  & $~0.05$  & $~0.94$  & $-0.71$  & $-0.62$ \\
 \end{tabular}
 \label{tab:corr_dimu}
 \end{center}
 \end{table}
 \clearpage
  \subsection{ $e \mu$ events}
\begin{figure}[htb]
 \begin{center}
 \vspace{-0.2in}
 \leavevmode
 \epsfxsize \textwidth
 \epsffile{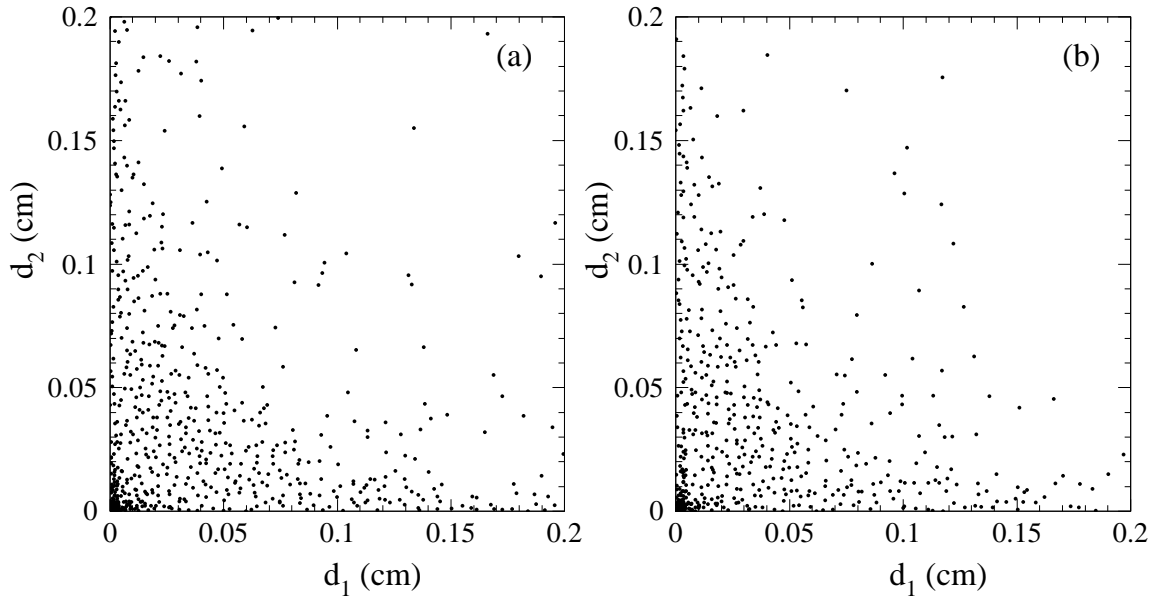}
 \caption[]{Two dimensional impact parameter distributions for (a) OS and
            (b) LS $e\mu$ events.}
 \label{fig:fig_6}
 \end{center}
 \end{figure}

\begin{figure}[htb]
 \begin{center}
 \vspace{-0.2in}
 \leavevmode 
 \epsfxsize \textwidth
 \epsffile{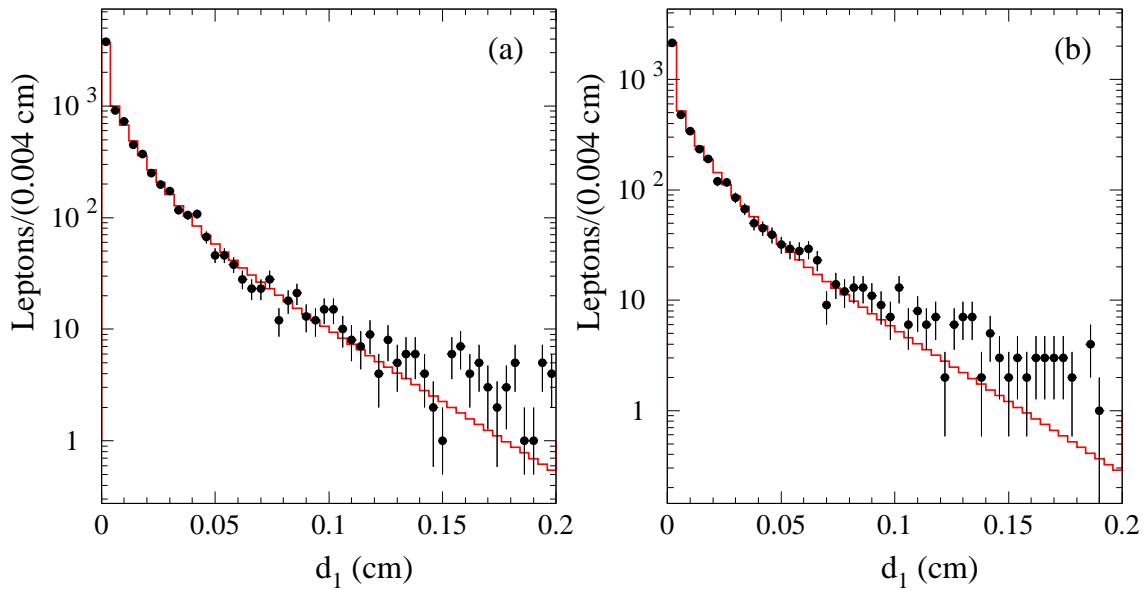}
 \caption[]{The projection of the impact parameter distribution of (a) OS 
            and (b) LS $e\mu$ pairs onto one of the two axis is compared 
            to the fit.}
 \label{fig:fig_7}
 \end{center}
 \end{figure}

   Figure~\ref{fig:fig_6} shows the observed two-dimensional impact parameter 
   distributions for OS and LS $e \mu$ pairs. There are 7802 OS and 4331 LS
   $e\mu$ events~\footnote{
   Since lepton tracks are reconstructed requiring at least two hits in the
   SVX detector close to the beam pipe, the number of electrons due to
   unidentified photon conversion is negligible (no larger than three).}.
   When all the likelihood terms are used to fit the data, the best fit, 
   as expected, returns $CC = 0 \pm 80$ LS events. However, while the fit
   finds an appreciable $BP$ component, it returns $CP = 0 \pm 130$ in both
   LS and OS events. As in the case of dimuon events, the fit gets blocked
   at the lower limits when the $CC(LS)$  and $CP$ parameters are bound to be
   positive, and we  exclude these terms in the fit likelihood. The fit result
   is shown in Table~\ref{tab:tab_2} and the parameter correlation matrix is
   listed in Table~\ref{tab:corr_emu}. The best fit returns $-\ln L =2481$. As
   for dimuon events, the probability of the  $-\ln L$ value returned by the 
   fit is determined by fitting Monte Carlo pseudoexperiments. We find that
   62\% of the fits to the pseudoexperiments return $-\ln L$ values equal or
   larger than 2481. For a comparison of the data and the fit result,
   projections of the two-dimensional impact parameter distributions are shown
   in Fig.~\ref{fig:fig_7}. Since the fit appears to underestimate the data 
   for $d_1 \geq$ 0.1 cm, we have fitted the data excluding points at impact
   parameters larger than 0.1 cm; this fit returns a result identical to that
   of the standard fit.

   Using Table~\ref{tab:tab_2} one derives that the ratio of LS to OS dileptons
   due to $b\bar{b}$ production is $R= 0.560 \pm 0.024$. 
\newpage
 \begin{table}[htb]
 \begin{center}
 \caption[]{Number of events attributed to the different sources by the fit
            to OS and LS $e \mu$ pairs. The errors correspond to a 0.5 change
            of the $-\ln L$.}
 \begin{tabular}{lcc}
  Component  &        OS      &        LS      \\
   $BB$      & $5099 \pm 138$ & $2852 \pm 90~$ \\
   $CC$      & $1126 \pm 162$ &   ~~~~0~~~     \\
   $PP$      & $~906 \pm 60~$ & $~875 \pm 52~$ \\
   $BP$      & $~536 \pm 107$ & $~529 \pm 102$ \\
   $CP$      &   ~~~~0~~~     &    ~~~~0~~~    \\
 \end{tabular}
 \label{tab:tab_2}
 \end{center}
 \end{table}
 \begin{table}[htb]
 \begin{center}
 \caption[]{Parameter correlation coefficient returned by the fit listed
            in Table~\ref{tab:tab_2}.}
 \begin{tabular}{lcccccc}
  Component  & $BB(OS)$ & $CC(OS)$ & $PP(OS)$ & $BP(OS)$ & $BB(LS)$ &$PP(LS)$\\
   $CC(OS)$  & $-0.63$  &          &          &          &          &        \\
   $PP(OS)$  & $~0.38$  & $-0.37$  &          &          &          &        \\
   $BP(OS)$  & $-0.23$  & $-0.33$  & $-0.43$  &          &          &        \\
   $BB(LS)$  & $~0.12$  & $~0.29$  & $~0.18$  & $-0.67$  &          &        \\
   $PP(LS)$  & $~0.31$  & $-0.14$  & $~0.76$  & $-0.56$  & $~0.23$  &        \\
   $BP(LS)$  & $-0.23$  & $-0.29$  & $-0.45$  & $~0.95$  & $-0.70$  & $-0.59$\\
 \end{tabular}
 \label{tab:corr_emu}
 \end{center}
 \end{table}
  \section{Average $B^0\bar{B}^0$ mixing probability}
  \label{sec:ss-chib}
   The average  $B^0\bar{B}^0$ mixing probability is defined as
   \begin{eqnarray*}
     \bar{\chi}= \frac{ \Gamma(B^0 \rightarrow \bar{B}^0 \rightarrow l^+ X )}
                      { \Gamma( B \rightarrow l^{\pm} X)} 
   \end{eqnarray*}
   where the numerator includes $B^0_d $ and $B^0_s$ mesons and the
   denominator includes all $B$ hadrons. In absence of mixing, the double
   semileptonic decay of a $B\bar{B}$ pair results in an OS lepton pair;
   when one of the mesons undergoes mixing a LS lepton pair is produced.
   The mixing probability $\bar{\chi}$ can therefore be inferred from $R$,
   the ratio of LS to OS dileptons due to $b\bar{b}$ production.

   The sequential decays of $b$-hadrons also contribute to $R$. The fraction
   of leptons from sequential decays, $f_l$, is evaluated using the simulation.
   Using simulated dimuon events, we find $f_{\mu}=0.123$ with a 12\%
   uncertainty~\footnote{
   Technically this fraction accounts also for the 0.4\% fraction of events
   which contain more than two hadrons with heavy flavor.}. 
   As for the study of Ref.~\cite{yale}, the uncertainty on $f_{\mu}$ 
   comes from the uncertainty of the relative branching ratios of $b$ and $c$
   semileptonic decays ($\pm 11$\%) and the uncertainty of the detector 
   acceptance for sequential leptons with respect to that for leptons from
   direct decays ($\pm 6$\%). Using the $e \mu$ simulation, we derive 
   $f_e=0.060$ and $f_{\mu}=0.142$ with a $\pm 12$\% systematic uncertainty.

   The ratio $R$ is related to the time-integrated mixing probability in the
   following way:
   \begin{eqnarray*}
   R=\frac{ f [\bar{\chi}^2+(1-\bar{\chi})^2]+2\bar{\chi}(1-\bar{\chi})(1-f)} 
          {(1-f)[\bar{\chi}^2+(1-\bar{\chi})^2]+2\bar{\chi}(1-\bar{\chi}) f }
   \end{eqnarray*}
   where $f=2 f_{\mu} (1-f_{\mu})=0.2157 \pm 0.0226\; ({\rm syst.})$ for 
   dimuon events and $f= f_e +f_{\mu} - 2 f_e f_{\mu}=0.1850 \pm 0.0204\;
   ({\rm syst.})$  for $e\mu$ events. Systematic errors due to other sources
   are negligible with respect to that arising from the $f$ uncertainty, and
   are neglected (see Sec.~\ref{sec:ss-cross}).

   From the observed values of $R$, we derive the following mixing
   probabilities:
  \begin{center}
  \begin{tabular}{ll}
   $\bar{\chi} = 0.136 \pm 0.009\; ({\rm stat.})\pm 0.014\; ({\rm syst.})$ & 
     for dimuon events \\
   $\bar{\chi} = 0.165 \pm 0.011\; ({\rm stat.})\pm 0.011\; ({\rm syst.})$ &
     for $e \mu$ events\\ 
  \end{tabular}
  \end{center}
   Since we use events containing two and only two leptons, the results from
   the dimuon and $e \mu$ data sets are statistically independent. Therefore,
   we combine the two results and derive an average mixing probability
   $\bar{\chi} = 0.152 \pm 0.007\; ({\rm stat.})\pm 0.011\;
   ({\rm syst.})$~\footnote{
   The systematic error quoted in Ref.~\cite{yale} ($\pm 0.016$)
   is larger to account for the fact that the $BP$ and $CC$ terms are
   not fitted independently.}.

   This value of the mixing probability agrees with all previous results
   from $p\bar{p}$ colliders

   $\bar{\chi}=0.157\pm 0.020\; ({\rm stat.})\pm 0.032\; ({\rm syst.})$
   (UA1~\cite{ua1})

   $\bar{\chi}=0.176\pm 0.031\; ({\rm stat. + syst.})\pm 0.032\; ({\rm model})$
   (CDF~\cite{emumix})

   $\bar{\chi}=0.131\pm 0.020\; ({\rm stat.})\pm 0.016\; ({\rm syst.})$
   (CDF~\cite{yale})\\
   but is significantly larger than the world average 
   $\bar{\chi}= 0.118 \pm 0.005$~\cite{pdg}, which is dominated by the LEP
   measurements at the $Z$-pole~\footnote{
   The world average assumes that the fractions $f_d$ and $f_s$ at the 
   Tevatron are equal to those at the $Z$-pole.}.
   Since our result is statistically very different from the world average, we
   have investigated the error behaviour beyond one $\sigma$. For an 8 unit
   increase of the $-\ln L$ value ($4~\sigma$ uncertainty), the errors of the
   $BB(OS)$ and $BB(LS)$ terms returned by the fit increase by a factor of
   four, and we derive a $4~\sigma$ statistical error of 0.029 for the combined
   value of $\bar{\chi}$.
  \section{Cross-checks of the result and study of additional
           systematic effects}
  \label{sec:ss-cross}
   In this section, we first perform several cross-checks of the $\bar{\chi}$
   result, and then investigate its sensitivity to the modeling of the 
   production and weak decay of heavy quarks. In subsection~A we verify that
   the ratio of the number of lepton pairs due to $c\bar{c}$ production to 
   that due to $b\bar{b}$ production returned by the various fits is 
   consistent with the theoretical expectation. Subsection~B compares our
   result to the previous CDF measurement, which used a subset of the data 
   available for this analysis. Subsection~B also verifies that the 
   $\bar{\chi}$ result is not affected by the small cosmic ray background 
   present in the dimuon data sample. Subsection~C shows that the $\bar{\chi}$
   result is not affected by the fact that we have excluded the $CP$ component
   in the fit likelihood. Subsections~D,~E and~F explore the dependence of our
   result on the mixture of the different $b$ and $c$ hadrons, on the ratio
   of $b\bar{b}$ to $c\bar{c}$ production cross section, and on the transverse
   momentum distribution of hadrons with heavy flavor predicted by the QCD
   simulation. In analogous measurements, these effects are usually not 
   considered since they are hard to quantify and to implement consistently 
   into the QCD generator. We investigate them either by changing the heavy
   flavor composition of the data with proper kinematical selections, or with
   reasonable modifications of the simulation prediction. Finally, subsections
   ~G and~H verify the templates used to separate the contribution of 
   semileptonic decays of heavy flavor from that of leptons due to 
   misidentified hadrons or prompt sources as the Drell-Yan process. We show 
   that all above effects change our result by a very small fraction of the 
   quoted statistical and systematic errors. We report changes in R when the 
   sequential fraction $f_l$ is not affected by the particular study, and 
   also changes in $\bar{\chi}$ when $f_l$ is affected; a summary of the 
   different results is presented in subsection~I. 
  \subsection{Ratio of the $c\bar{c}$ to $b\bar{b}$ production}
  \label{sec:ss-bcrat}
   The difference between the  $\bar{\chi}$ measurements at the Tevatron and
   LEP may not require an explanation in terms of new physics; however, if
   we entertain the hypothesis~\cite{berger} that the enhancement of the
   $b\bar{b}$ cross section at the Tevatron with respect to the NLO
   prediction may be caused by pair production of light gluinos decaying to a
   bottom quark and a bottom squark, which in turn produces an apparent
   increase of $\bar{\chi}$ with respect to LEP, then the ratio of the
   $c\bar{c}$ to $b\bar{b}$ cross sections should be approximately a factor 
   of two smaller than what is predicted by the Standard Model. Therefore, 
   it is of interest to compare the ratio of the numbers of leptons due to 
   $c\bar{c}$ and $b\bar{b}$ production in the data and the simulation. 

   The dimuon fit in Table~\ref{tab:tab_1} returns a ratio 
   $\frac{CC}{BB}=0.15 \pm 0.02\; ({\rm stat.})$. In the simulation, 
   this ratio is $0.18 \pm 0.02\; ({\rm stat.})$.

   The fit to $e \mu$ data in Table~\ref{tab:tab_2} returns a ratio
   $\frac{CC}{BB}=0.14 \pm 0.02\; ({\rm stat.})$. In the simulation,
   the ratio is 0.12 with a negligible statistical error.
 
   As shown in Ref.~\cite{scaqua}, which studies events with jets 
   corresponding to partons with transverse momentum larger than 20 $\gevc$,
   the {\sc herwig} generator predicts heavy flavor cross sections which are
   approximately a factor of two larger than the NLO calculation~\cite{mnr}
   and models correctly the $c\bar{c}$ and $b\bar{b}$ cross section observed
   at the Tevatron. However, muons in the present analysis correspond to
   partons with $p_T \geq 6.5\; \gevc$ (electrons to partons with $p_T \geq 9\;
   \gevc$). A priori, there is no guarantee that {\sc herwig} still does a
   good job in predicting the ratio $\frac {CC}{BB}$ also in this data set
   which corresponds to a hard scattering with smaller transverse momenta
   (the inclusive $b\bar{b}$ and $c\bar{c}$ cross sections are approximately
   a factor of 40 larger in this data set than in the jet data studied in
   Ref.~\cite{scaqua}). We cross-check the ratio of the $c\bar{c}$ to 
   $b\bar{b}$ parton-level cross sections evaluated with {\sc herwig} with two
   different NLO Monte Carlo calculations. In {\sc herwig}, the ratio of the 
   $c\bar{c}$ to $b\bar{b}$ cross sections for producing both heavy quarks
   with $|\eta| \leq 1$ and transverse momentum large enough to produce an
   electron with $E_T \geq 5$ GeV and a muon with $p_T \geq 3\; \gevc$ is
   1.37. In the {\sc mnr} calculation~\cite{mnr}, this ratio is found to be
   1.39, while the {\sc cascade} Monte Carlo generator~\cite{cascade} predicts
   a value of 1.39~\cite{jung}.

   We conclude that the ratio of dileptons due to $c\bar{c}$ production to 
   that due to $b\bar{b}$ production at the Tevatron is consistent with the
   prediction of the presently available Monte Carlo generators.
  \subsection{Cosmic ray background in dimuon events and comparison
  	     with the previous CDF result}
  \label{sec:ss-comp}
   The previous CDF measurement of $\bar{\chi}$~\cite{yale} uses a subset
   (17.4 pb$^{-1}$) of the dimuon sample (105 pb$^{-1}$) collected by CDF
   and used in the present analysis. There are minor differences in the data
   selection. In the present analysis we exclude dimuons with impact
   parameters $d_1 +d_2 \geq 0.2$ cm to reduce the impact of the cosmic ray
   background, and we exclude the $\Upsilon$ invariant mass region which has 
   a negligible fraction of heavy flavor contribution.

   To study our sensitivity to the cosmic ray background we have performed
   a fit to the data which includes dimuons with $d_1 +d_2 \geq 0.2$ cm.
   This fits returns a ratio $R = 0.533 \pm 0.018$ (the standard fit yields
   $R = 0.537 \pm 0.018$). We conclude that the small cosmic ray background
   does not affect the fit result.

   In order to compare with the result in Ref.~\cite{yale} we fit the data
   including the $\Upsilon$ mass region. Because of the slightly different 
   selection, the total number of events in the present analysis, 35265, is
   24\% larger than the number of events selected in Ref.~\cite{yale}
   (4750 events) multiplied by the ratio of the relative luminosities.
   The fit which includes this mass region is shown in Table~\ref{tab:tab_1_a}.
   The fit returns a total of $18737 \pm 275$ dimuon events due to $b\bar{b}$
   production. Consistently, this number is 25\% larger than the number of  
   dimuon events attributed in Ref.~\cite{yale} to $b\bar{b}$ production 
   ($2471 \pm 104$ events) multiplied by the ratio of the relative 
   luminosities. This fit that includes the $\Upsilon$ mass region yields 
   $R = 0.535 \pm 0.017\; ({\rm stat.})$, which compares well to the result
   of our standard fit and the value $R=0.502 \pm 0.041\; ({\rm stat.)}$ in
   Ref.~\cite{yale}. 
 \begin{table}[htb]
 \begin{center}
 \caption[]{Number of events attributed to the different sources of dimuons
            by the fit to OS and LS dimuons including the invariant mass
            region between 9.2 and 10.5 GeV/$c^2$.}
 \begin{tabular}{lcc}
 Component  &         OS      &          LS    \\
  $BB$      & $12202 \pm 237$ & $6535 \pm 139$ \\
  $CC$      & $~2849 \pm 388$ &   ~~~~0~~~     \\
  $PP$      & $~7601 \pm 189$ & $2173 \pm 94~$ \\
  $BP$      & $~1662 \pm 175$ & $1658 \pm 167$ \\
  $CP$      &    ~~~~~0~~~    &    ~~~~0~~~    \\
 \end{tabular}
 \label{tab:tab_1_a}
 \end{center}
 \end{table}
  \subsection{Effect of neglecting the $CP$ component in the 
	      likelihood function}
  \label{sec:ss-nocp}
   In order to estimate correctly the uncertainties of the $b\bar{b}$ and
   $c\bar{c}$ contributions returned by the fit, we had to set to zero the
   $CP$ component, which is expected to be of the same size of the $BP$
   component~$^9$. We have performed a number of pseudo-experiments of
   approximately the same size and composition as the data. In each 
   pseudo-experiment, the impact parameters of the dileptons contributed by 
   a given component are extracted from the corresponding two-dimensional
   template used to fit the data. Each pseudo-experiment has been fitted as 
   the data, and the result of 125 pseudo-experiments is shown in
   Table~\ref{tab:tab_sys_1}. In 15\% of the pseudo-experiments, the $CP$
   value returned by the fit is so close to zero that the fit gets blocked 
   at the lower limit; as for the data, the $CP$ term has to be ignored in
   the likelihood in order to estimate correctly the uncertainty of the
   $BB$ term.
 \begin{table}[htb]
 \begin{center}
 \caption[]{Number of generated and fitted events in 125 pseudo-experiments.
            We list the average and the rms spread of the values returned by
            the fits.}
 \begin{tabular}{lcc}
  Component  & generated  &     fitted     \\
   $BB$      & $8000$     & $7998 \pm 247$ \\
   $CC$      & $4000$     & $3991 \pm 544$ \\
   $PP$      & $4000$     & $3999 \pm 348$ \\
   $BP$      & $1200$     & $1204 \pm 505$ \\
   $CP$      & $1200$     & $1196 \pm 812$ \\
 \end{tabular}
 \label{tab:tab_sys_1}
 \end{center}
 \end{table}
   We have further investigated the sensitivity of the $R$ result to the 
   value of the $CP$ component returned by the fit by constraining it to be
   equal to the $BP$ contribution within the statistical error. The fit 
   results are shown in Table~\ref{tab:tab_sys_2} for dimuon events and in  
   Table~\ref{tab:tab_sys_3} for $e \mu$ events. These fits return 
   $R=0.533 \pm 0.016$ (the standard fit returns $R=0.537 \pm 0.018$) for 
   dimuon events and $R=0.559 \pm 0.023$ (the standard fit returns
   $R=0.560 \pm 0.024$) for $e \mu$ events.
 \begin{table}[htb]
 \begin{center}
 \caption[]{Number of events attributed to the different sources of dimuons by
            the fit to OS and LS dimuons with $d_1+d_2 \leq 0.2$ cm. 
            The errors correspond to a 0.5 change of $- \ln L$.}
 \begin{tabular}{lcc}
  Component  &         OS      &        LS      \\
   $BB$      & $10691 \pm 232$ & $5695 \pm 134$ \\
   $CC$      & $~2203 \pm 404$ &    ~~~~0~~~    \\
   $PP$      & $~3328 \pm 166$ & $1536 \pm 122$ \\
   $BP$      & $~1009 \pm 130$ & $1001 \pm 126$ \\
   $CP$      & $~~878 \pm 122$ & $~869 \pm 117$ \\
 \end{tabular}
 \label{tab:tab_sys_2}
 \end{center}
 \end{table}
 \begin{table}[htb]
 \begin{center}
 \caption[]{Number of events attributed to the different sources by the fit
            to OS and LS $e \mu$ events. The errors correspond to a 0.5 change
            of $- \ln L$.}
 \begin{tabular}{lcc}
  Component  &        OS      &        LS     \\
   $BB$      & $5171 \pm 134$ & $2892 \pm 92$ \\
   $CC$      & $1083 \pm 162$ &   ~~~~0~~     \\
   $PP$      & $~798 \pm 70~$ & $~767 \pm 64$ \\
   $BP$      & $~312 \pm 63~$ & $~308 \pm 60$ \\
   $CP$      & $~300 \pm 61~$ & $~293 \pm 58$ \\
 \end{tabular}
 \label{tab:tab_sys_3}
 \end{center}
 \end{table}
 \subsection{Sensitivity to the $b$ and $c$ lifetime}
 \label{sec:ss-life}
   The impact parameter distribution of leptons from $b$ and $c$ decays has
   some dependence on the lifetime uncertainty. We have varied the average
   $b$-hadron lifetime in the simulation by $\pm 10 $\% and refit the data 
   with the resulting templates in order to investigate which effect might
   have the possibility that the relative fractions of different $b$-hadrons
   in the simulation are grossly different from the data. The fractions of 
   the $BB$ components, which are returned by the fit, change by approximately
   $\pm 9$\% for both OS and LS dileptons; however, the ratio $R$ changes 
   by less than 0.2\%.  

   Since $c\bar{c}$ events contribute only to OS events, we have studied the
   sensitivity of the fit to the impact parameter template for $c$
   semileptonic decays. We have constructed impact parameter templates by 
   varying in the simulation the relative ratio of $D^{\pm}$ to $D^0$ mesons
   by $\pm 30$\%~\footnote{The lifetime is $c \tau=315\; \mu$m for the
   $D^{\pm}$ meson and $c \tau=123\; \mu$m for the  $D^{0}$ meson.}. 
   The $CC$ component in OS dileptons returned by the fit changes by 
   approximately $\pm 10$\%. In the fit, this change is mostly compensated by
   the $BP$ component, and the $BB$ contribution to OS dilepton changes 
   by less than $\pm 0.1$\%.
  \subsection{Sensitivity to the $c\bar{c}$ contribution}
  \label{sec:ss-charm}
   The $c\bar{c}$ production contributes only OS dileptons. The value of $R$ 
   returned by the fit can be affected by a poor modeling of this contribution.
   We investigate this possibility by analyzing a data sample with a smaller
   fraction of $c\bar{c}$ contribution. According to the {\sc herwig}
   generator program, and also to the {\sc mnr} Monte Carlo program~\cite{mnr},
   the ratio of the $c\bar{c}$ to $b\bar{b}$ cross sections for producing
   both heavy flavor partons with $|\eta| \leq 1$ and transverse momenta 
   larger than 9 GeV/c is 1 while in the simulation of the standard $e \mu$
   data set is 1.37.

   This kinematical situation is modeled by selecting muons, as well as 
   electrons, with $p_T \geq 5$ GeV/c. We derive from the simulation of this
   data set new impact parameter templates for $b$- and $c$-hadron decays.
   The fit result is shown in Table~\ref{tab:tab_4}. The fit yields
   $R = 0.524 \pm 0.034$. In this case, the fractions of sequential decays
   are $f_e=0.060$, $f_{\mu}=0.092$, and $f=0.1410\pm 0.0158\; ({\rm syst.})$.
   It follows that $\bar{\chi}= 0.170\pm 0.015\; ({\rm stat.})\pm 0.007\;
   ({\rm syst.})$, in agreement with the result of the standard fit 
   $\bar{\chi} = 0.165\pm 0.011\; ({\rm stat.})\pm 0.011 ({\rm syst.})$.  
 \begin{table}[htb]
 \begin{center}
 \caption[] {Number of events attributed to the different sources by the fit
             to OS and LS $e \mu$ events in which both leptons have 
             $p_T \geq 5$ GeV/c. The errors correspond to a 0.5 change 
	     of $- \ln L$.}
 \begin{tabular}{lcc}
  Component  &        OS     &        LS     \\
   $BB$      & $2113 \pm 86$ & $1107 \pm 57$ \\
   $CC$      & $~421 \pm 98$ &     ~~~~0~~   \\
   $PP$      & $~265 \pm 36$ & $~249 \pm 31$ \\
   $BP$      & $~163 \pm 68$ & $~159 \pm 65$ \\
   $CP$      &   ~~~~0~~     &     ~~~~0~~   \\
 \end{tabular}
 \label{tab:tab_4}
 \end{center}
 \end{table}
  \subsection{Sensitivity to the modeling of the kinematics}
  \label{sec:ss-kin}
   Because we select leptons above a certain $p_T$ threshold, the impact
   parameter templates for leptons from semileptonic decays of heavy flavors
   have some dependence on the modeling of the $p_T$ distribution of the
   parent hadron with heavy flavor~\footnote{In the extreme case of a lepton
   with $p_T$ close to the 5 $\gevc$ threshold, parent hadrons with a 5 GeV
   transverse energy produce leptons with zero impact parameter.}.
   The modeling of the $p_T$ distribution of the parent hadron with heavy
   flavor can be affected by a wrong estimate of the relative contribution of
   processes of order $\alpha_s^{2}$ and $\alpha_s^{3}$, or by an incorrect
   modeling of the hadronization of heavy quarks~\footnote
   {In the simulation partons arising from $\alpha_s^{2}$ diagrams are slightly
   stiffer than those contributed by $\alpha_s^{3}$ diagrams.}. 
   In the next two subsections, we investigate the sensitivity of our result
   to these effects.
  \subsubsection{Dileptons with $\delta \phi \geq 2.4 $}
  \label{sec:ss-dphi}
   According to the  simulation, the fractional contribution of $b\bar{b}$ 
   and $c\bar{c}$ direct production (LO term) increases with increasing
   $\delta \phi$, the azimuthal opening angle between the two leptons. Using
   dileptons with $\delta \phi \geq 2.4$ rad, the number of simulated events
   due to $b\bar{b}$ and $c\bar{c}$ production is reduced by 64\% and 66\%,
   respectively. At the same time, the fraction of direct production in 
   $b\bar{b}$ events increases from 71\% to 84\% and the fraction of direct
   production in $c\bar{c}$ events increases from 66\% to 76\%. 

   Using this selection, the data consist of 4872 OS and 2745 LS dileptons.
   The result of the fit to these events using standard templates is shown 
   in Table~\ref{tab:tab_3}. We derive $R = 0.576 \pm 0.032$,
   in good agreement with the standard fit result $R=0.560 \pm 0.024$.
 \begin{table}[htb]
 \begin{center}
 \caption[] {Number of events attributed to the different sources by the fit
             to OS and LS $e \mu$ events with $\delta \phi \geq 2.4 $. 
             The errors correspond to a 0.5 change of $-\ln L$.}
 \begin{tabular}{lcc}
  Component  &        OS      &        LS     \\
   $BB$      & $3255 \pm 110$ & $1874 \pm 75$ \\
   $CC$      & $~688 \pm 129$ &   ~~~~0~~     \\
   $PP$      & $~534 \pm 47~$ & $~513 \pm 41$ \\
   $BP$      & $~314 \pm 88~$ & $~310 \pm 84$ \\
   $CP$      &    ~~~~0~~~    &   ~~~~0~~     \\
 \end{tabular}
 \label{tab:tab_3}
 \end{center}
 \end{table}
  \subsubsection{Dependence on the $p_T$ spectrum of the parent hadron
                 with heavy flavor}
  \label{sec:ss-pt}
\begin{figure}
 \begin{center}
 \vspace{-0.3in}
 \leavevmode
 \epsffile{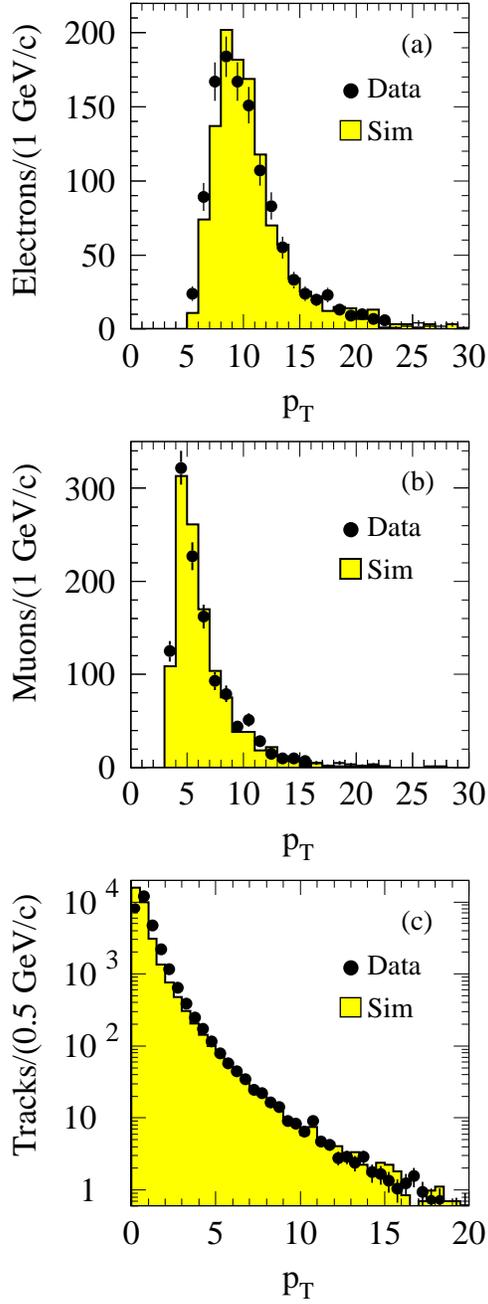}
 \caption[]{Comparison of the transverse momentum distributions of electrons
            (a) and muons (b) in the data and in the heavy flavor simulation. 
            The bottom plot (c) shows the transverse momentum distribution of 
            all other tracks in $e \mu$ events. Data and simulation are
            normalized to the same number of events.}
 \label{fig:fig_datsim}
 \end{center}
 \end{figure}

   As shown by Fig.~21 of Ref.~\cite{suj} and Figures~7 and~8 of 
   Ref.~\cite{scaqua}, our simulation models quite well the hadronization of 
   $b$ and $c$ quarks with transverse energy larger than 20 GeV. As shown in
   Fig.~\ref{fig:fig_datsim}, the simulation also models correctly the 
   lepton transverse momentum distributions in the $e \mu$ data. Because the
   lepton distribution depends on the $p_T$ distribution of the parent parton
   and its fragmentation function, we use a comparison between data and
   simulation to evaluate their global uncertainty. A fit of the lepton $p_T$
   spectra with the simulated shapes weighted with the function $p_T^{\alpha}$,
   where $\alpha$ is a free fit parameter, returns $\alpha=0.003 \pm 0.023$.
   In the simulation, such changes of lepton $p_T$ distributions can be 
   modeled by reweighting the $p_T$ distribution of the parent parton with 
   the function $p_T^{\pm \beta=0.5}$. Fits to the $e \mu$ data using 
   templates constructed with these modified simulations return 
   $R=0.557 \pm 0.024$ for $\beta=0.05$ and $R=0.559 \pm 0.024$ for 
   $\beta=-0.05$ (the result of the standard fit is $0.560 \pm 0.023$).
  \subsection{Dependence on the modeling of the impact parameter distributions}
  \label{sec:ss-4svx}
   For tracks in a jet, the impact parameter resolution in the data is
   slightly larger  than in the parametrized {\sc qfl} detector simulation 
   which has in input the SVX-hit resolution of the data~\cite{topxs}. This
   is believed to be due to the  probability of reconstructing a track with
   spurious SVX hits, which in the data is larger than in the simulation
   because the SVX occupancy in the data is also larger. In JET~20 
   data~\footnote {Events collected with a trigger that requires at least 
   one jet with $E_T \geq $20 GeV.}, 
   the transverse  energy deposited by charged tracks in a cone of radius 0.2
   in the $\eta- \phi$ space around the axis of a lepton contained in a jet is
   $\simeq$ 18 GeV. For the events used in this analysis, the transverse 
   energy deposited by charged tracks in a cone of radius 0.2 around each
   lepton is $\simeq$ 0.8 GeV; in this case, the transverse momentum 
   distribution of all charged tracks in the dilepton events, plotted in  
   Fig.~\ref{fig:fig_datsim}(c), is also well modeled by the simulation.

   To further investigate the sensitivity to spurious SVX hits, we have 
   repeated our study by using only leptons with 4 SVX hits; we also require
   that at least two of the hits are not shared with other tracks. We also 
   make use of new templates for prompt leptons, and leptons from $b$- and 
   $c$-hadron decays constructed using this track selection.

   With this selection, the dimuon data consist of 9822 OS and 4785 SS pairs.
   Table~\ref{tab:tab_5} lists the result of the fit to dimuon events passing
   this selection. The fit yields $R = 0.548 \pm 0.025$, in good agreement
   with the result of the standard fit $R = 0.537 \pm 0.018$.

\begin{figure}[htb]
 \begin{center}
 \vspace{-3.in}
 \leavevmode
 \epsfxsize \textwidth
 \epsffile{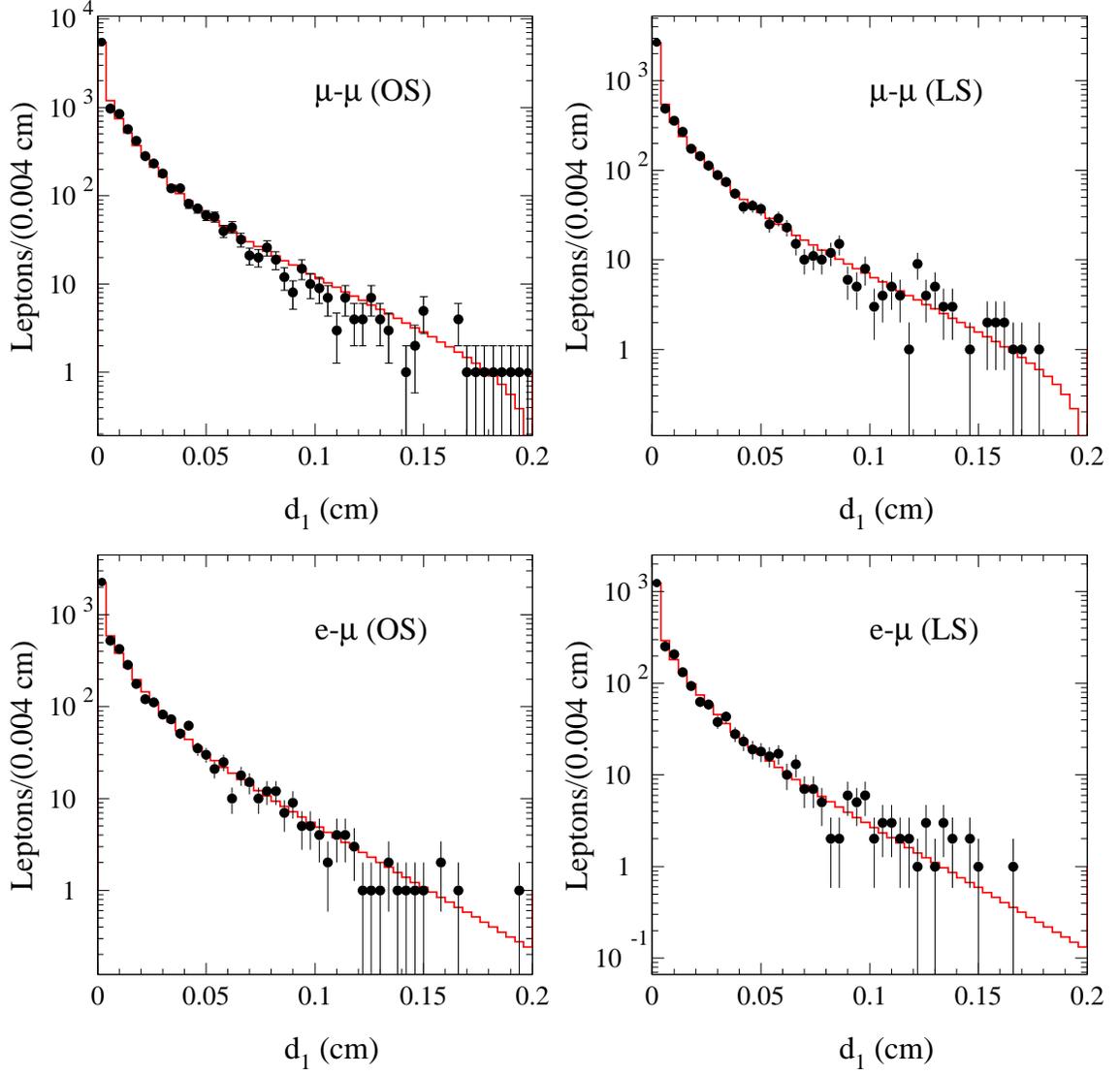}
 \caption[]{The projection of the impact parameter distributions in the data
            ($\bullet$) is compared to the the fit result (dashed histograms).}
 \label{fig:fit_4hits}
 \end{center}
 \end{figure}

   The $e \mu$ data consist of 4465 OS and 2355 SS pairs with 4 SVX hits.
   Table~\ref{tab:tab_6} lists the fit result. The fit yields 
   $R = 0.559 \pm 0.029$, in good agreement with the result of the standard
   fit $R = 0.560 \pm 0.024$. For a comparison of the data and the fit results,
   projections of the two-dimensional impact parameter distributions are shown
   in Fig.~\ref{fig:fit_4hits}. The combined result yields an average mixing
   parameter
   $\bar{\chi}=0.154\pm 0.009\; ({\rm stat.})\pm 0.011\; ({\rm syst.})$, 
   to be compared to the standard fit result 
   $\bar{\chi}=0.152\pm 0.007\; ({\rm stat.})\pm 0.011\; ({\rm syst.})$.
\newpage
 \begin{table}[htb]
 \begin{center}
 \caption[]{Number of events attributed to the different sources by the fit 
	    to OS and LS dimuons with 4 SVX hits and $d_1+d_2 \leq 0.2$ cm. 
            The errors correspond to a 0.5 change of the $-\ln L$.}
 \begin{tabular}{lcc}
  Component  &        OS      &        LS      \\
   $BB$      & $4990 \pm 150$ & $2735 \pm 90~$ \\
   $CC$      & $1818 \pm 245$ &    ~~~~0~~~    \\
   $PP$      & $2237 \pm 112$ & $1289 \pm 63~$ \\
   $BP$      & $~740 \pm 110$ & $~743 \pm 106$ \\
   $CP$      &   ~~~~0~~~     &    ~~~~0~~~    \\
 \end{tabular}
 \label{tab:tab_5}
 \end{center}
 \end{table}
 \begin{table}[htb]
 \begin{center}
 \caption[]{Number of events attributed to the different sources by the fit 
            to OS and LS $e \mu$ events with 4 SVX hits. The errors correspond
	    to a 0.5 change of the $-\ln L$.} 
 \begin{tabular}{lcc}
  Component  &        OS      &        LS     \\
   $BB$      & $2768 \pm 99~$ & $1547 \pm 66$ \\
   $CC$      & $~831 \pm 121$ &    ~~~~0~~    \\
   $PP$      & $~575 \pm 44~$ & $~552 \pm 37$ \\
   $BP$      & $~266 \pm 76~$ & $~264 \pm 73$ \\
   $CP$      &    ~~~~0~~~    &    ~~~~0~~    \\
 \end{tabular}
 \label{tab:tab_6}
 \end{center}
 \end{table}
 \clearpage
  \subsection{Leptons faked by tracks from hadronic decays of hadrons
              with heavy flavor}
  \label{sec:ss-jetfak}
   In the standard fit to the  data, we have approximated the impact parameter
   distribution of fake leptons with that of leptons from prompt sources. The
   fits return a $BP$ component which is 15\% (dimuon events) and 10\% 
   ($e \mu$ events) of the $BB$ component. According to the simulation, only
   7.5\% of the events due to the $BP$ component contain
 a second hadron with heavy flavor 
   which decays hadronically;  in these events, less than 50\%
   of the tracks, which are fake-lepton candidates, arise from the decay of
   the heavy flavored hadron; in addition, 80\%
   of the lepton faked by tracks from hadronic decays of heavy flavors carry
   a charge with the same sign of that of the parent heavy flavor quark. 
   Therefore, one estimates that the effect of this approximation
   on $R$ is of the order of $10^{-3}$~\footnote{
   This is supported by the fact the $CC$ component in LS dilepton, which can
   only be contributed by leptons faked by tracks from hadronic decays of 
   charmed hadrons, is found negligible by our fit with a 1 $\sigma$ upper 
   limit of 1.6\% of the $CC$ contribution to OS dileptons.}.

\begin{figure}
 \begin{center}
 \vspace{-1.in}
 \leavevmode
 \epsffile{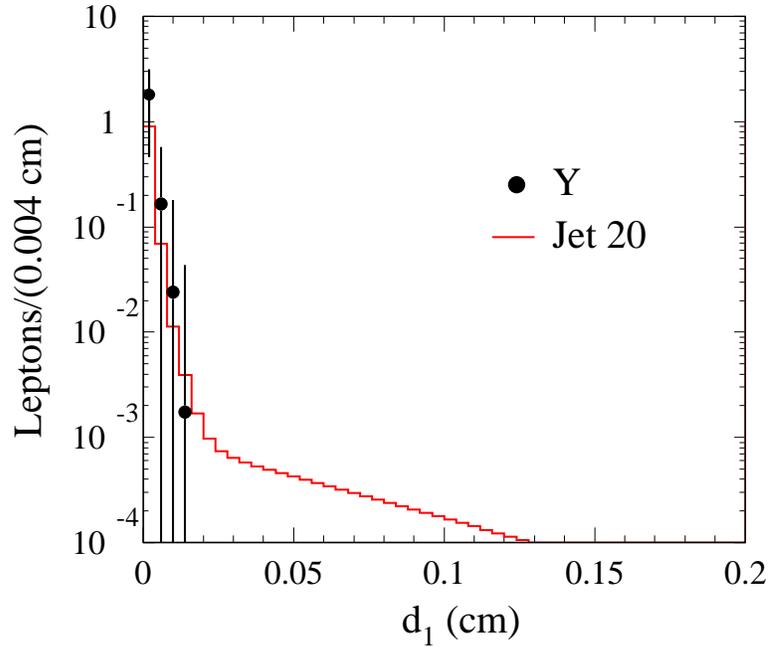}
 \caption[]{Comparison of the impact parameter distributions of lepton 
	    candidate tracks in JET~20 data and of leptons coming from 
            $\Upsilon(1S)$ decays.}
 \label{fig:fig_8}
 \end{center}
 \end{figure}

   We cross-check our conclusion by modeling fake leptons with new templates,
   called $F$ (instead of $P$), derived in a sample with a comparable 
   contamination of hadrons with heavy flavor. This sample consists of events
   containing a jet with $E_T \geq 20$ GeV. As shown by the study in 
   Ref.~\cite{topxs}, JET~20 data contain a 9.5\% fraction of heavy flavor.
   After removing events in which jets contain a soft lepton (SLT tag)
   or a displaced secondary vertex (SECVTX tag), the contamination of heavy 
   flavor is 7.1\% (comparable to the fraction of heavy flavor with hadronic
   decay contributing to the $BF$ and $CF$ components). The new template is
   constructed by using all tracks with $p_T \geq 3\; \gevc$ and pointing to
   the CMUP fiducial volume. Figure~\ref{fig:fig_8} compares the new template
   to the one derived using prompt muons.
   Tables~\ref{tab:tab_7} and~\ref{tab:tab_8} list the results of the fits to
   dilepton events with 4 SVX hits when using templates which account for the
   heavy flavor contribution to fake leptons. The fits return
   $R = 0.570 \pm 0.027$ for dimuon events, and $R = 0.562 \pm 0.034$ for 
   $e \mu$ events. The combined result yields an average mixing probability
   $\bar{\chi}=0.159\pm 0.010\; ({\rm stat.})\pm 0.011\; ({\rm syst.})$ 
   to be compared to the standard fit result 
   $\bar{\chi}=0.154\pm 0.009\; ({\rm stat.})\pm 0.011\; ({\rm syst.})$.
\newpage
 \begin{table}[htb]
 \begin{center}
 \caption[]{Number of events attributed to the different sources of dimuons by
            the fit to OS and LS dimuons with 4 SVX hits and 
            $d_1+d_2 \leq 0.2$ cm. Fake leptons for the $BF$ and $CF$
            components are modeled with a template derived in JET~20 data.}
 \begin{tabular}{lcc}
  Component  &        OS      &        LS      \\
   $BB$      & $4781 \pm 150$ & $2723 \pm 90~$ \\
   $CC$      & $2207 \pm 222$ &    ~~~~0~~~    \\
   $PP$      & $2018 \pm 111$ & $1251 \pm 64~$ \\
   $BF$      & $~787 \pm 108$ & $~796 \pm 104$ \\
   $CF$      &    ~~~~0~~~    &   ~~~~~0~~~    \\
 \end{tabular}
 \label{tab:tab_7}
 \end{center}
 \end{table}
 \begin{table}[htb]
 \begin{center}
 \caption[]{Number of events attributed to the different sources of dimuons by
            the fit to OS and LS $e \mu$ with 4 SVX hits. Fake leptons for the
            $BF$ and $CF$ components are modeled with a template derived in
            JET~20 data.}
 \begin{tabular}{lcc}
  Component  &        OS      &        LS     \\
   $BB$      & $2743 \pm 103$ & $1541 \pm 68$ \\
   $CC$      & $~857 \pm 118$ &     ~~~~0~~   \\
   $PP$      & $~586 \pm 45~$ & $~566 \pm 38$ \\
   $BF$      & $~257 \pm 76~$ & $~256 \pm 73$ \\
   $CF$      &    ~~~~0~~~    &     ~~~~0~~   \\
 \end{tabular}
 \label{tab:tab_8}
 \end{center}
 \end{table}
  \subsection{Summary of the cross-checks}
   Table~\ref{tab:tab_sum} lists the $\bar{\chi}$ values resulting from the 
   different cross-checks presented in this section. All $\bar{\chi}$ 
   measurements are consistent with the main result presented in 
   Sec.~\ref{sec:ss-chib}.
 \begin{table}[htb]
 \begin{center}
 \caption[]{Summary of the cross-checks presented in Sec.~\ref{sec:ss-cross}.
            The $\bar{\chi}$ error is statistical only.}
 \begin{tabular}{clc}
  Data set            & fit type                  & $\bar{\chi}$      \\
 $\mu \mu + e\mu$     & standard                  & $0.152 \pm 0.007$ \\
 $\mu \mu + e\mu$     & $BP = CP$  (Sec.~\ref{sec:ss-nocp})
					          & $0.151 \pm 0.007$ \\
 $\mu \mu + e\mu$     & 4 SVX hits (Sec.~\ref{sec:ss-4svx})  
						  & $0.154 \pm 0.009$ \\
 $\mu \mu + e\mu$     & 4 SVX hits, JET~20 fakes (Sec.~\ref{sec:ss-jetfak}) 
                                                  & $0.159 \pm 0.010$ \\
 $ e \mu$             & standard                  & $0.165 \pm 0.011$ \\
 $ e \mu$             & $\Delta \phi \geq 2.4$ rad (Sec.~\ref{sec:ss-dphi}) 
 						  & $0.173 \pm 0.015$ \\
 $ e \mu$             & $p^{lepton}_T \geq 5\; \gevc$ (Sec.~\ref{sec:ss-charm})
					          & $0.170 \pm 0.015$ \\
 $ e \mu$             & $\beta = +0.05$ (Sec.~\ref{sec:ss-pt})
					          & $0.164 \pm 0.011$ \\
 $ e \mu$             & $\beta = -0.05$ (Sec.~\ref{sec:ss-pt})
					          & $0.165 \pm 0.011$ \\
 \end{tabular}
 \label{tab:tab_sum}
 \end{center}
 \end{table}
  \clearpage
  \section{Conclusions}
  \label{sec:ss-concl}
   Using samples of $\mu \mu$ and $e \mu$ pairs collected with the CDF
   experiment during the $1992-1995$ run of the Tevatron collider, we have
   performed a high precision measurement of $\bar{\chi}$, the time integrated
   mixing probability of $b$-flavored hadrons produced at the Tevatron.
   Our measurement, $\bar{\chi}=0.152\pm 0.007\; ({\rm stat.})\pm 0.011\;
   ({\rm syst.})$, confirms the trend of all previous results from $p\bar{p}$
   colliders, and is significantly larger than the world average  
   $\bar{\chi} = 0.118 \pm 0.005$, which is dominated by the LEP measurements
   at the $Z$-pole.
  \section{Acknowledgments}
   We thank the Fermilab staff and the technical staff of the participating
   Institutions for their contributions. This work was supported by the
   U.S.~Department of Energy and National Science Foundation; the Istituto
   Nazionale di Fisica Nucleare; the Ministry of Education, Culture, Sports,
   Science and Technology of Japan; the Natural Science and Engineering 
   Research Council of Canada; the National Science Council of the Republic
   of China; the Swiss National Science Foundation; the A.P.~Sloan Foundation;
   the Bundesministerium f\"{u}r Bildung und Forschung; the Korea Science and
   Engineering Foundation (KoSEF); the Korea Research Foundation; and the
   Comision Interministerial de Ciencia y Tecnologia, Spain.

 \end{document}